\documentclass{elsart}
\usepackage{graphicx,amssymb}
\journal{Chemical Physics Letters}
\begin{document}
\begin{frontmatter}
\title{Characterization of doping levels in heteronuclear, gas-phase, van der Waals clusters and 
their energy absorption from an intense optical field}
\author[tifr]{J. Jha},
\author[barc]{P. Sharma},
\author[barc]{V. Nataraju},
\author[barc]{R. K. Vatsa},
\author[tifr]{D. Mathur\corauthref{cor}},
\corauth[cor]{Corresponding author.}
\ead{atmol1@tifr.res.in}
and
\author[tifr]{M. Krishnamurthy}
\address[tifr]{Tata Institute of Fundamental Research, 1 Homi Bhabha Road,
Mumbai 400 005, India}
\address[barc]{Bhabha Atomic Research Centre, Trombay, Mumbai 400 085, India}

\begin{abstract}
A simple mass spectrometric method has been developed to quantify dopant levels in heteronuclear 
clusters in the gas phase. The method is demonstrated with reference to quantification of the 
water content in supersonic beams of water-doped argon clusters. Such doped clusters have assumed 
much importance in the context of recently-reported doping-induced enhancement in the emission of 
energetic charged particles and photons upon their interaction with intense laser pulses. We have 
also measured the energy that a doped cluster absorbs from the optical field; we find that energy 
absorption increases with increasing level of doping. The oft-used linear model of energy 
absorption is found to be quantitatively inadequate.
\end{abstract}
\end{frontmatter}

\section{Introduction}

	Clusters have unique properties compared with atoms, molecules and bulk matter. This is 
manifested most prominently in the interaction of clusters with intense laser fields 
\cite{RostReview06,smirnov,mathur1}. In recent experiments, irradiation by short, intense, 
infrared (800 nm wavelength) laser pulses with large van der Waals clusters comprising a few 
thousand to several hundred thousand atoms has been extensively investigated. One of the 
motivations for such studies has been the unusually efficient manner in which large, gas-phase 
clusters absorb energy from the optical field that they are irradiated by. Energy absorption in 
excess of 90\% has been reported for rare-gas clusters exposed to laser intensity in the range of 
10$^{16}$ W cm$^{-2}$ \cite{ditabs}. One consequence of this energy absorption is the ejection of 
a large number of electrons from the cluster. The multiply-ionized cluster subsequent undergoes 
Coulomb explosion, giving rise to the ejection of energetic highly-charged ions \cite{asym}, 
electrons \cite{SpringatePRA}, x-ray photons \cite{Mocek}, and the occurrence of nuclear fusion 
\cite{ditfusion}.  Such emissions are of obvious importance from the viewpoint of developing 
tabletop sources of energetic charged particles and photons. It is, therefore, not surprising that 
considerable effort has been expended in recent years to vary parameters like laser polarization 
\cite{ourprl}, pulse width \cite{ion}, and chirp \cite{zamith} so as to facilitate some measure of 
control over the brightness and yield of the emitted charged particles and photons. 

An alternative possibility of controlling cluster ionization dynamics is by {\em altering} not the 
laser parameters but the constituents of the cluster target. This approach was initially suggested 
by Castleman and coworkers \cite{CastlemanPapers} in the context of very small clusters (of size 
less than $\sim$100 atoms); they introduced hydrogen iodide molecules into a stream of Ar gas in a 
supersonic jet expansion cluster source. Charge transfer processes in such few-atom clusters were 
postulated to lead to a change in the charge states of Ar atoms due to the presence of easily 
ionizable HI. It was also shown that when easily ionizable HI is present in Kr cluster expansion, 
higher Kr charge states were obtained \cite{CastlemanPapers}. Very recently, an extension of this 
approach has been demonstrated to control and enhance the x-ray emission from nano-cluster plasmas 
that are formed upon irradiation by intense femtosecond-duration laser pulses: when very large 
argon clusters (comprising tens of thousands of atoms) are doped with water molecules the 
time-integrated yield of Ar K x-ray emission has recently been shown to be enhanced nearly 
twelve-fold in comparison to pure Ar clusters under otherwise identical experimental conditions 
\cite{xRayJPhysBL}. It appears established that by using low ionization energy dopants, it becomes 
possible to significantly enhance both the high-energy ion yield and the charge states that are 
obtained from heteronuclear clusters \cite{IonsAPL06}. 

The doping of Ar clusters with H$_2$O molecules (which have lower ionization energy than Ar atoms) 
alters the ionization dynamics of the clusters: the relative ease of ionizing the water molecules 
increases the number of electrons that are ionized from the cluster constituents in the initial 
temporal evolution of the optical pulse. This larger number of ionized electrons brings about an 
enhanced contribution to the optical field driven electron-impact ionization. In the laser-cluster 
interaction the main ionization mechanisms are (1) optical field ionization (OFI), which includes 
tunnel ionization and barrier suppression ionization \cite{RostReview06,smirnov}, (2) collisional 
ionization \cite{ditmirePRA53}, and (3) charge-enhanced ionization in the combined field of ions 
and the laser radiation \cite{RosePetruckXrayPRA}. A simple OFI ionization model \cite{smirnov} 
estimate shows that the intensity threshold for H$_2$O is around three times lower than that for 
Ar. With our experimental parameters, Ar$_{n}$ achieves 100\% ionization when each Ar is converted 
into Ar$^+$ at an intensity of 2.5$\times$10$^{14}$ W cm$^{-2}$ \cite{ditmirePRA53}. For a 100 fs 
FWHM Gaussian pulse with peak laser intensity of $\sim$10$^{16}$ W cm$^{-2}$, conversion to Ar$^+$ 
occurs at time ($t$-117) fs, where $t$ denotes the time at which the pulse intensity peaks. The 
appearance intensity for a particular charge state of a given atom or ion is $\sim I_{p}^{4}$ 
\cite{smirnov}, where I$_{p}$ is the ionization energy of the species involved. Therefore, one 
electron from each water molecule will appear at an intensity of 1$\times$10$^{14}$ W cm$^{-2}$, 
which occurs at time ($t$-129) fs. Note that this time is 18 fs earlier than the time at which the 
first ionization of pure argon occurred.  Furthermore, the enhanced ionization mechanism 
\cite{bandrauk} that is so important for molecular ionization in strong fields, would lead to 
further increase of the ionization rate for H$_2$O. It is, therefore, seen that the electron 
density is enhanced  in (Ar-H$_2$O)$_n$ as compared to pure Ar$_n$. To eject the second electron 
from a water molecule would require another 16 eV 
\cite{SmithMueller,MathurPhysRep1,MathurPhysRep2}; this ionization step would occur at time 
($t$-114) fs. By the time the laser intensity reaches $\sim$2.6$\times$10$^{14}$ W cm$^{-2}$, 
argon atoms are only singly ionized while water dopants are already doubly ionized. The ionized 
electrons may still be confined to the cluster as a whole due to Coulombic forces  \cite{Jortner}.  
In the intensity regime 10$^{16}$ W cm$^{-2}$ and for cluster sizes of more than a few nm, a 
substantial fraction of electrons are trapped in the potential well of the cluster \cite{Jortner}. 
Energy from the incident optical field couples to the cluster-plasma through these trapped 
electrons. These quasi-free electrons which are delocalized  within the cluster, oscillate under 
the combined influence of the rapidly-increasing intra-cluster electrostatic field and the intense 
optical field \cite{ldh}, and give rise to the optical field driven electron impact ionization we 
referred to above.  The inherent time-dependent nature of the entire process makes the temporal 
profile of the electron density a most critical parameter for deposition of optical energy into 
the cluster plasma. The consequent steeper rise in the electron density profile within the cluster 
thus results in enhancement of the propensity of high charge states being formed \cite{IonsAPL06} 
as well as in significant enhancement of the yield of x-rays \cite{xRayJPhysBL}. 

A most important lacuna that has plagued all recent reports on doped clusters (and the spectacular 
effects that doping produces in the context of irradiation by short-duration, intense optical 
fields) is lack of quantitative information on the level of doping that is achieved. For instance, 
the obvious question that arises from experiments on water-doped Ar clusters concerns 
quantification of the percentage of water that is present in the heteronuclear cluster. The 
vapor-pressure of water at room temperature is about 20 Torr. At atmospheric pressure, a typical 
mixture of argon-water is expected to contain about 2.6\% of water. In all hitherto-reported 
experiments on water-doped clusters, argon gas was bubbled through water to accomplish the doping. 
Therefore, on the basis of simple partial pressure arguments pertaining to a water-argon mixture, 
the upper limit on the percentage of water in the doped clusters cannot be more than $\sim$2.6\%. 
At high values of stagnation pressure of argon, at which laser-cluster experiments are typically 
conducted \cite{xRayJPhysBL,IonsAPL06}, the proportion of water is likely to be even less than 
this notional limit as factors like dynamic momentum transfer and sticking coefficients make the 
overall situation very complex. These factors warrant a much more careful doping estimate than one 
that is simply based on equilibrium vapor-pressure type of arguments. 

The quantification task remains challenging as a typical set-up for laser-cluster experiments 
would not be equipped to analyze doping levels of a few percent. We report here a relatively 
simple method based on mass spectrometry that helps to quantify the fraction of water in a doped 
heteronuclear cluster in an experimental environment that closely resembles that utilized in 
typical laser-cluster experiments. We demonstrate our method with reference to the amount of water 
dopant in an argon cluster, and we show that the enhancement in laser energy absorption increases 
with the doping level, in consonance with recently reported enhancements of x-ray and charged 
particle energies and yields under similar water-doping circumstances. 

\section{Experimental Method} 
 
Our experimental technique is based upon a quadrupole mass-spectrometric (QMS) analysis of species 
of various charge/mass values in a seeding situation of the type that pertains to doped cluster 
formation. A schematic view of our experimental setup is shown in Fig. 1. Argon gas jet is bubbled 
through triply-distilled water at various values of backing pressure. A buffer container is used 
to avoid spluttering of water droplets inside the experimental chamber. The Ar-water mixture is 
ionized by a beam of electrons of mean energy of 70 eV, with an energy spread of $\sim$1 eV. The 
ionized gas mixture undergoes subsequent charge/mass analysis by the QMS. The background pressure 
of the ionization chamber is maintained at $\sim$10$^{-7}$ Torr, with a typical working pressure 
of $\sim$10$^{-6}$ Torr with full gas load. The pressure gauge head (Pfeiffer, PKR-251) to monitor 
vacuum in the QMS chamber and the leak valve (Pfeiffer, EVR-116) connected to the same chamber are 
electronically controlled through a RVC 300 controller to maintain a constant working pressure 
inside the experimental chamber. 

	In order to recreate the experimental conditions that are typical of laser-cluster interaction 
experiments, argon is bubbled through water at different values of backing pressures. The 
high-pressure argon-water mixture is then passed through a T-tube, one port of which leaks a part 
of the mixed gas to the atmosphere. The leak to the atmosphere is adjusted so that a maximum of 1 
bar of gas pressure is behind the leak valve into the apparatus. The mixture is then allowed to 
effuse into the ionization chamber in a controlled way through the RV 300 controller. As the 
effusion rate is inversely proportional to the square root of mass of the species, water will 
effuse 1.49 times faster than argon. This fact is taken into account while estimating the water 
content in the mixture. 

The QMS generates mass spectra that allow us to determine the area under the curve of each m/q 
peak. Proper consideration of the ionization cross-sections of argon and water allows 
quantification of the proportion of each species. 

For measurements of laser absorption by doped and undoped clusters, we used a supersonic, 
solenoid-operated Parker General Valve that was fitted with a 500 $\mu$m nozzle of 45$^\circ$ half 
expansion angle \cite{xRayJPhysBL}. The cluster sizes were estimated using the empirical Hagena 
parameter \cite{smirnov,ditmirePRA53}. Within the experimental parameters used in our 
measurements, the cluster size varies from about 350 atoms at 2 bar to about 15000 atoms at 10 
bar. A Ti:Sapphire laser system that delivers up to 55 mJ (800 nm, wavelength) of 100 fs pulses at 
10 Hz repetition rate was used. Laser pulses of energy up to 10 mJ/pulse were focused in the 
supersonic jet expansion by a 25 cm plano-convex lens to attain intensities in the 10$^{15}$ W 
cm$^{-2}$ range. The scattered and transmitted 800 nm light was measured using a combination of 
bandpass filters and energy-calibrated, fast photodiodes placed at various locations.

It is relevant to mention that while our method is very simple, it is not expected to be as 
accurate as the standard gravimetric method used by gas-phase reaction kineticists 
\cite{Gravimetry1,Gravimetry2}. Our primary aim is to estimate the water content in mixed Ar-water 
clusters. In typical laser-cluster experiments, a major source of uncertainty lies in cluster size 
determination. The standard Hagena method \cite{smirnov,ditmirePRA53} of estimation of cluster 
size gives the value of mean cluster size irrespective of the distribution of cluster sizes around 
the mean value. (A detailed discussion of cluster size distributions has been presented elsewhere 
, see \cite{LogNormal}, and references therein.) The Hagena estimate has an inherent error of 
about 30\%. With this limitation on cluster size information, the accuracy of the method we 
present here suffices. Moreover, the simplicity of the method enables a nearly ``on-line" 
determination of doping levels to be made, a facet that normal gravimetric methods would not 
allow. We also note that mass flow controllers used in the gravimetric method usually do not work 
reliably in the pressure range much above 3 bar while doped cluster formation experiments usually 
require considerably higher pressure values. 

\section{Results and Discussion}

Figure 2 shows the percentage of water that is present in the mixture as a function of stagnation 
pressure, as estimated from the QMS spectra. Particular care has been taken to get a ``true" 
sample of the argon-water mixture: the proportion of water in the mixture is estimated as follows. 
At an incident electron energy of 70 eV, we take the electron impact ionization cross-section for 
the first ionization of neutral Ar to be 2.67 (in units of 10$^{-16}$ cm$^{2}$), while that for 
the second ionization is 0.146 \cite{straub}. In case of water, the respective cross-sections for 
ionization to H$_2$O$^+$ and OH$^+$ are 1.62 and 0.95 \cite{orient}. The inset in Fig. 2 shows a 
typical QMS trace when a mixture of argon and water sample was obtained upon bubbling argon 
through water at 5 bar backing pressure and, thereafter, being ionized by 70 eV electrons. 
Individual m/q species are indicated by arrows.  

The proportion of water is seen to initially increase with backing pressure, starting from about 
2.5\% doping level at 1 bar backing pressure to a maximum of about 7\% doping level at nearly 6 
bar backing pressure. Thereafter, the doping level falls to about 2\% at 8 bar. It is relevant to 
emphasize here that at backing pressure of 1 bar, the experimental measurement yields a value of 
2.5\% as water doping level; this value is, within the associated error bars, quite consistent 
with the value that is deduced using simple partial pressure arguments, as discussed in the 
Introduction. However, with increase of the argon backing pressure, a larger amount of argon (as 
density is proportional to the pressure) becomes available to ``push" water molecules into forming 
a mixture. But, as pressure is further increased, the bulk flow velocity of argon atoms increases 
and, as argon gas bubbles through a finite column of water, the time that an argon atom makes 
contact with water reduces, as does the likelihood of momentum transfer. Consequently, there is 
competition between the inverse of time ($\sim$velocity) that argon spends in the vicinity of 
water and the number density of argon atoms. Beyond a certain value of pressure, in our case 
around 6 bar, the higher flow velocity of argon atoms becomes more important than the increment in 
argon atom density. Consequently, the amount of water that is carried by the swiftly flowing argon 
atoms is reduced. In order to reproduce the experimental conditions used in the typical doped 
cluster experiments \cite{xRayJPhysBL,IonsAPL06}, the flow conditions which include the stagnation 
pressure, diameter of the gas line, height of water column, etc. were kept as close to those in 
\cite{xRayJPhysBL,IonsAPL06} as possible. For this reason alone we do not need to take into 
account the change in temperature of water as argon at high pressure was allowed to pass through 
it: our measurements do not need to make an {\em absolute} determination of the number of water 
molecules. After measurement of relative H$_2$O and Ar ion signals at each value of stagnation 
pressure we replenished the water sample and allowed sufficient time between subsequent 
measurements to allow for temperature equilibration. 

Figure 3 shows a typical mass spectral trace when argon was directly fed into our ionization 
chamber, bypassing the water reservoir. This measurement provides an estimate for the background 
water present in the experimental chamber and gas-feed lines and provides an aid to properly 
estimating the actual doping level in the argon-water mixture. As can be estimated from Fig. 3, 
the relative yield of H$_2$O$^+$ and OH$^+$ is 1.1, while the ionization cross-sections would 
indicate this to be about 1.7. There are likely to be other sources of OH$^+$ in the ionization 
chamber, but the measured ratio is properly accounted for in all data analysis.

In order to correlate the extent of doping with the enhancements that have been recently reported 
in the interactions of intense, short-duration laser light with doped clusters we also carried out 
laser absorption measurements. Figure 4 shows some typical results of absorption measurements that 
we made for pure and doped Ar clusters as a function of stagnation pressure,  using a fixed laser 
intensity of $\sim$10$^{15}$ W cm$^{-2}$. We collected the scattered 800 nm laser light along the 
laser polarization, and at an angle of 90$^o$ to it, and also the transmitted light using a 
combination of bandpass filters and energy-calibrated, fast photodiodes placed at various 
locations. Given the relevant differential scattering cross-section \cite{Kerker,Zweiback12}, 
these measurements suffice to give a good estimate of the absorbed laser energy. Our measurements 
were confined to relatively low values of stagnation pressure but, nevertheless, they suffice to 
clearly demonstrate that absorption of laser energy is significantly more in the case of doped 
cluster. It is seen that within the parameter ranges of the experiments, the difference in 
absorption follows the trend that is indicated by the doping levels that can be quantified from 
data shown in Fig. 2. In the following we present a possible scenario which could explain the 
enhancement in the absorption of laser energy in case of doped clusters.

Clusters sizes of the type of relevance to the present study are typically a few nm, which are 
much smaller than the wavelength (800 nm) of the laser used in such experiments. Therefore, a 
cluster will experience an optical field that is uniform across its dimensions. Consequently, the 
typical Rayleigh type of dipole model ought to suffice in order to account for most of the 
observed light scattering phenomena \cite{Kerker} and, indeed, this is found to be the case for a 
large number of absorption and scattering type of laser-cluster experiments 
\cite{RostReview06,smirnov,Zweiback12}. In this model, the absorption cross-section for a cluster 
of radius $r$ is given by
\begin{equation}
\sigma_{abs} = 4 \pi k r^{3} Im\lbrack{(\epsilon-1)/(\epsilon+2)}\rbrack,
\end{equation}
where $k$ = $\omega$/c is the propagation constant of the laser (see \cite{Zweiback12}, and 
references therein, $\omega$ is the angular frequency of the laser light and c is its speed in 
vacuum. $\epsilon$ is the dielectric constant of the cluster medium and is given within the simple 
Drude model \cite{ditmirePRA53, Zweiback12} by
\begin{equation}
\epsilon = 1 - {\omega_{p}^{2}\over{\omega^{2}}}\times{1\over{1+i\nu/\omega}},
\end{equation}
where $\omega_{p}$ denotes the plasma frequency and $\nu$ is the electron-ion collision frequency.
For a 1 keV plasma exposed to laser intensity of $\sim$10$^{15}$ W cm$^{-2}$, the electron-ion 
collision frequency for a typical argon cluster turns out to be 0.21 fs$^{-1}$ \cite{Liu1}, 
considering the typical ionic charge to be 8 and electron density in the argon cluster to be 160 
nm$^{-3}$. These values are reasonable as the charge state of 8+ for Ar$_{40,000}$ clusters has 
been directly demonstrated by means of two-dimensional time-of-flight mass spectrometry 
\cite{ourprl,ChargeDist}. Using these values, it is possible to deduce an estimate of 
$Im\lbrack{(\epsilon-1)/(\epsilon+2)}\rbrack$ to be 3.3$\times$10$^{-3}$ for homonuclear clusters 
of Ar atoms. Taking the percentage of water dopant in our heteronuclear clusters to be around 
10\%, as has been demonstrated in our data, and considering the parameters used in very recent 
experiments on ion ejection from doped Ar-H$_2$O clusters \cite{IonsAPL06}, the estimate for 
$Im\lbrack{(\epsilon-1)/(\epsilon+2)}\rbrack$ turns out to be about 23\% larger for water-doped 
clusters. However, as is clear from our experimental data, the actual enhancement of absorption 
between undoped and doped clusters is more than a factor of two over the entire range of 
stagnation pressures (cluster sizes) in our experiment. This discrepancy points to the need to 
develop models that will adequately take account of the fact that the optical field induces 
effects in clusters that are essentially nonlinear in nature. The model that is used here, and by 
others \cite{RostReview06,smirnov} takes cognizance of only linear effects.

We note that there is recent experimental evidence that doping does not entail a change in the 
size of the cluster \cite{xRayJPhysBL}.  The variation of the absorption with pressure that we 
observe is in consonance with the only earlier report \cite{ditabs} available in the literature.

In summary, we have presented a very simple technique to estimate the water content in water-doped 
argon clusters. Our techniques is applicable to many other doping situations in the context of 
heteronuclear clusters. Doped clusters have recently assumed much importance in the light of very 
significant enhancement in charged particle and photon emissions that ensue upon interactions with 
ultrashort, intense laser pulse. The estimated percent of water dopant in the mixed clusters has 
also been correlated with the laser absorption measurements that we have conducted. Doped clusters 
absorb laser energy much more efficiently. Our results show that quantification of doping levels 
is possible and this will open the way for more studies on the role of low ionization energy 
dopants in the strong-field dynamics of laser-cluster interactions. Our energy absorption 
measurements show that a significant enhancement occurs when argon clusters are doped with water. 
It is not possible to rationalize the extent of enhancement using a linear model of the type that 
is reported in the literature \cite{RostReview06,smirnov}. Our data brings to the fore the need to 
develop adequate nonlinear models that will properly account for energy absorption by the cluster 
from the optical field.

\newpage
\begin{figure}
\caption{\label{f1} Experimental setup for analysis of the argon-water mixture (see text). QMS: 
quadrupole mass spectrometer; CEM: channel electron multiplier detector; TMP: turbomolecular 
pump.}
\end{figure}

\begin{figure}
\caption{\label{f2} Percentage of water in the argon-water mixture as a function of stagnation 
pressure. The inset shows a typical mass spectral trace obtained when argon is bubbled through 
water at 5 bar backing pressure, species are indicated by arrows. The peak corresponding to the 
N$_2^+$ ion fragment is from the background nitrogen present in the experimental chamber at base 
pressure.} 
\end{figure}

\begin{figure}
\caption{\label{f3} A typical background mass spectral trace obtained when argon gas is directly 
ionized by 70 eV electrons. The peaks corresponding to various ionic fragments of water are from 
the background water vapor present in the chamber.} 
\end{figure}

\begin{figure}
\caption{\label{f4} Absorption of laser energy by clusters as a function of stagnation pressure 
for argon and doped Ar-H$_2$O clusters at a fixed incident laser intensity of 10$^{15}$ W 
cm$^{-2}$ and wavelength of 800 nm. The data were averaged over 1000 laser shots.}
\end{figure}

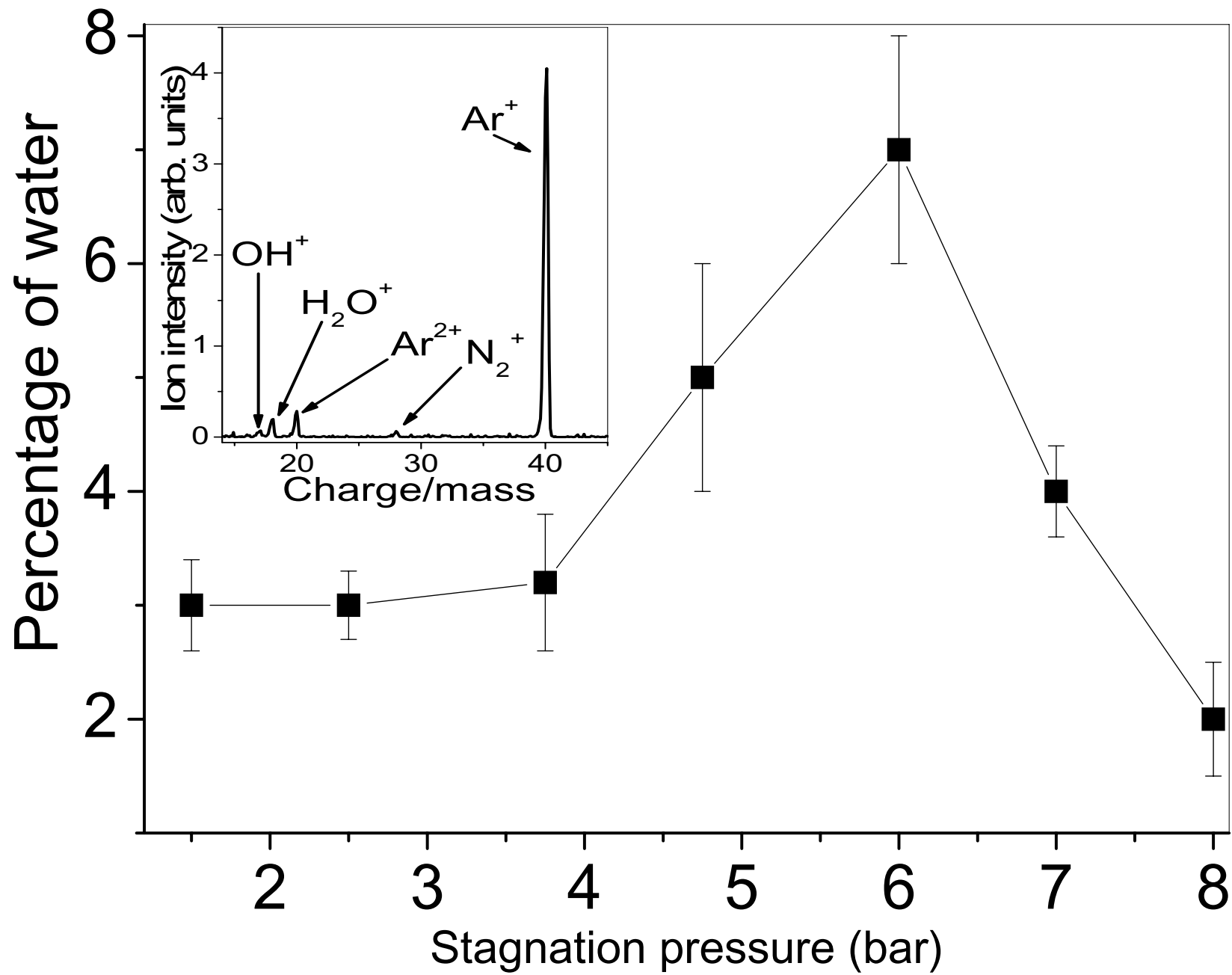
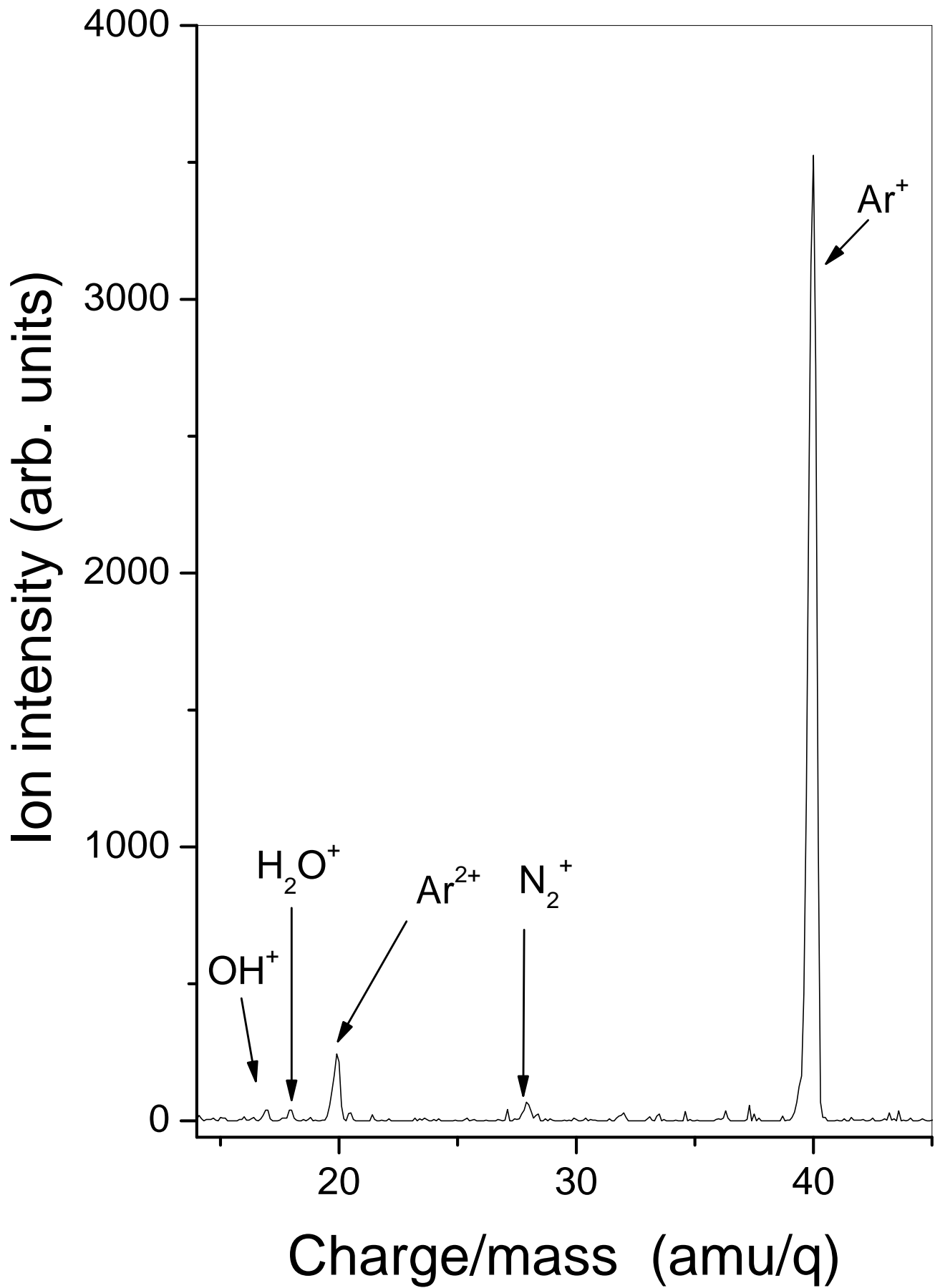
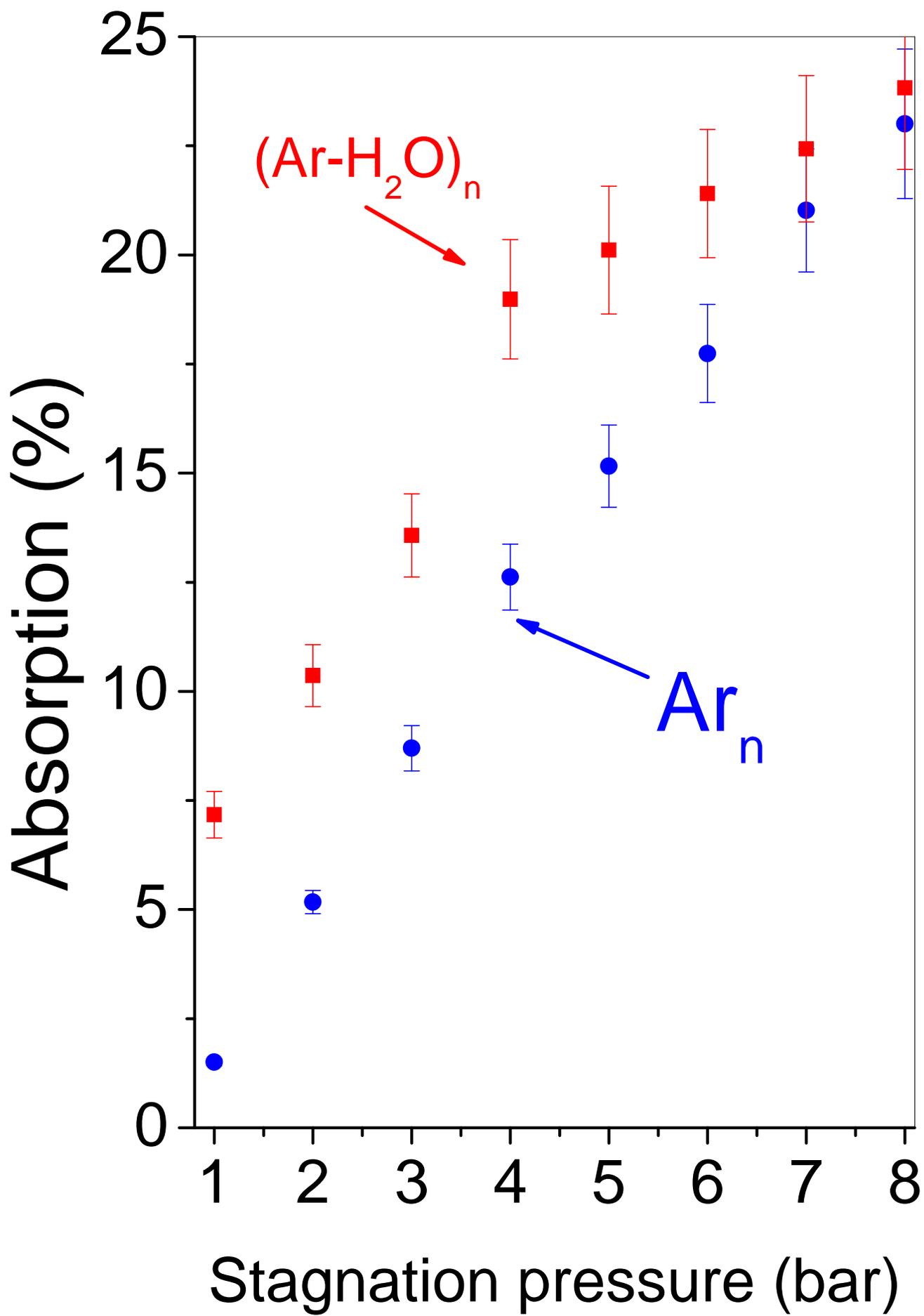
\end{document}